\documentstyle[12pt,fleqn,epsfig]{article}
\begin{document}
      
\begin{center}
{\bf INSTITUT~F\"{U}R~KERNPHYSIK,~UNIVERSIT\"{A}T~FRANKFURT}\\
D - 60486 Frankfurt, August--Euler--Strasse 6, Germany
\end{center}

\hfill IKF--HENPG/01--00

\vspace{1.5cm}
\begin{center}
{\Large \bf 
Baryon Number Conservation} 
\end{center}  
\begin{center}
{\Large \bf 
and
}
\end{center}  
\begin{center}
{\Large \bf 
Statistical Production of Antibaryons
}
\end{center}  

\vspace{1cm}

\begin{center}

Mark I. Gorenstein$^{a,}$\footnote{Permanent address: Bogolyubov Institute
for Theoretical Physics,
Kiev, Ukraine}$^,$\footnote{E--mail: goren@th.physik.uni-frankfurt.de},
Marek
Ga\'zdzicki$^{b,}$\footnote{E--mail: marek@ikf.physik.uni--frankfurt.de}
and Walter Greiner$^a$\footnote{E--mail: greiner@th.physik.uni-frankfurt.de}
\end{center}

\vspace{0.5cm}

$^a$ Institut f\"ur Theoretische Physik, Universit\"at  Frankfurt,
Germany

$^b$ Institut f\"ur Kernphysik, Universit\"at Frankfurt,
Germany

   

\vspace{1cm}

\begin{abstract}
\noindent
The statistical production of antibaryons is considered
within the canonical ensemble formulation.
We demonstrate that the antibaryon suppression
in small systems due to the exact
baryon number conservation is rather different in the baryon--free
($B=0$) and baryon--rich ($B\ge 2$) systems.
At constant values of temperature and baryon density 
in the baryon--rich systems
the density
of the produced antibaryons is only weakly dependent on the size
of the system. For realistic hadronization conditions
this dependence
appears to be close
to $B/(B+1)$ which is in agreement with
the preliminary data of the NA49 Collaboration 
for the $\bar{p}/\pi$ ratio in nucleus--nucleus collisions 
at the CERN SPS energies.
However,
a consistent picture of antibaryon production 
within the statistical hadronization model 
has not yet been achieved.
This is because the condition of constant hadronization temperature 
in the baryon--free systems leads to a contradiction with the data
on the $\bar{p}/\pi$ ratio in e$^+$+e$^-$ interactions.

\end{abstract}

\newpage
\noindent
{\bf 1. Introduction}

Among the first models of multiparticle
production in high energy  interactions were
statistical models \cite{Fe:50, Po:51, La:53}.
In the last decade a significant development of these
models and the extension of the area of their
applicability took place.
The main reason for this is a surprising success
of the statistical approach in reproducing new experimental data
on hadron multiplicities in 
nuclear (A+A) \cite{Go:99} and elementary (e$^+$+e$^-$, p+p,
p+$\bar{\rm{p}}$) collisions \cite{Be:96, Be:97}.
One of the important results of the analysis of hadron yield
systematics at high energies (SPS and higher)
done within the statistical models  
is the approximate independence
of the temperature parameter $T = 160\div 190$~MeV 
from the system size and collision energy \cite{temp}.
This result can be attributed to
the statistical character of the hadronization process.

The  statistical models are based on the key 
assumption that all microscopic states of the system 
allowed by the conservation laws are equally probable.
Calculations within these models are straightforward
when the mean number of particles of interest is large
and consequently it is enough to fulfill the conservation
laws in the average sense, i.e., for the  macroscopic state.
This is achieved  by the 
introduction of
parameters: the temperature $T$ 
and chemical potentials $\mu_i$, which control
correspondingly the average values of the system energy density and
its material content (e.g., baryon number, strangeness and electric charge). 
In this case (grand canonical ensemble, $g.c.e.$)
the mean particle multiplicities
are just proportional to the volume $V$ of the system.
The particle density and the ratio of the multiplicities of two
different particles often used for the comparison with the data
are volume independent.

This simple volume dependence is however not valid 
any more for a small system
in which the  mean particle multiplicity is low. In this case
(canonical ensemble, $c.e.$)
the material conservation laws 
should be imposed on each microscopic 
state of the system. 
This condition introduces a significant correlation 
between particles 
which carry conserved charges. 
The correlation reduces the effective
number of degrees of freedom and consequently leads to the 
$c.e.$ suppression of the 'charged' particle multiplicity when compared
with the result of the calculations done within $g.c.e.$
A 'neutral' particle (a particle which does not carry conserved
charges) does not feel the $c.e.$ suppression in the small
system, its mean multiplicity remains proportional to the
volume.
Therefore, a  different dependence on the system volume is
expected within statistical models for 'charged' and 'neutral'
particles. 
The magnitude of the $c.e.$ suppression of the 'charged'-'anticharged'
pair creation increases with the mass 
of the lightest hadron needed to compensate the 
particle charges. 
Therefore, a strong $c.e.$ suppression
may be expected for  antibaryon production as the mass of the  lightest
baryon is $m\cong 938$~MeV which is much larger than the value of the
temperature parameter found in the
hadronization models. 
The $c.e.$ suppression is still rather essential for
strange particle production (see, e.g., \cite{Ke:99}). 
  
The above expectation seems to be {\it violated} by the 
preliminary data on antiproton production presented
recently by the NA49 Collaboration \cite{Ve:99}. 
The $\bar{p}/\pi$ ratio is found to be approximately
the same for p+p interactions and central Pb+Pb
collisions at 158 A$\cdot$GeV ({\it antiproton scaling}).
Thus the hadronization volume increases but the
effect of the $c.e.$ suppression is not observed.
The $c.e.$ suppression can be expected in p+p interactions because of
the key difference between antiproton and pion:
the antiproton carries baryon number in addition to the
electric charge carried by both particles.
In order to compensate the electric charge of a produced particle
it is enough to create an additional charged pion.
As the pion mass is much smaller than the nucleon mass,
a significantly stronger $c.e.$ suppression is
expected for antiproton production than for the pion production,
consequently it should lead to a strong violation of the experimentally
observed scaling.
Thus the crucial question is whether the
{\it antiproton scaling} can be understood 
within the statistical
model of hadron production in which the condition of 
exact baryon number
conservation is imposed.
We note that the antiproton multiplicities in high energy collisions
were shown to approximately agree with the
predictions of statistical models \cite{Go:99,Be:96,Be:97}.   
However different versions of the models 
with different parameters were used to fit various
sets of data. 
Thus the question whether  consistent description of the
antiproton data in the statistical model is possible is
still opened.

The importance of the exact treatment of the material 
conservation laws within statistical models of strong
interactions
was first pointed out 
by Hagedorn \cite{Ha:71} (see also Refs.~\cite{Re:80,Ra:80}).
Subsequently a complete treatment has been developed
(see, e.g., \cite{Cl:97} and references therein) and applied
to analyze the hadron yields in elementary collisions
\cite{Be:96, Be:97}.
In this letter we  derive explicit analytical formulae to study the role
of the exact material conservation laws  
within the statistical model of hadronization and we use them to 
discuss
the {\it antiproton scaling} observed experimentally
in Pb+Pb
collisions at 158 A$\cdot$GeV \cite{Ve:99}.
We also discuss the data on the $\bar{p}/\pi$ ratio 
in e$^+$+e$^-$ interactions
\cite{pbar, pi} within the statistical hadronization model.

\vspace{0.5cm}
\noindent
{\bf 2. Model formulation}

Let us  consider
the system of baryons '$b$' and antibaryons '$a$'
with total baryon number $B$  
as the Boltzmann ideal gas in the volume $V$, at temperature $T$.
The {\it c.e.} partition function is
\begin{eqnarray}\label{partfunc}
Z(T,V,B)~&=&~\sum_{N_b^{(1)},N_a^{(1)}}^{{\infty}}
...\sum_{N_b^{(j)},N_a^{(j)}=0}^{\infty}...~~
\delta_K\left[B-\sum_j(N_b^{(j)}-N_a^{(j)})\right] \\
&\times &\prod _{j}~\frac{(\lambda_b^{(j)} z_j)^{N_b^{(j)}}}{N_b^{(j)}!}
\frac{(\lambda_a^{(j)} z_j)^{N_a^{(j)}}}{N_a^{(j)}!}~,\nonumber
\end{eqnarray}
where the index $j$ runs over all (non-strange) baryon states $N,\Delta, 
N^*,...$, and the single baryon (antibaryon) partition function reads 
\begin{eqnarray}\label{zfunc}
z_j~=~z_j(T,V)~&=&~\frac{g_j V}{(2\pi)^3}\int d^3k~ 
\exp[-(k^2+m_j^2)^{1/2}/T]~=~ \\
~&=&~ \frac{g_j 
V}{2\pi^2}~T~m^2_j~K_2(m_j/T)~\equiv~V~f_j(T)~.\nonumber 
\end{eqnarray}
The baryon  mass and the 
baryon degeneracy factor are denoted here by $m_j$ and $g_j$, respectively. 
Auxiliary
parameters $\lambda_b^{(j)}$ and $\lambda_a^{(j)}$
are introduced in order 
to calculate the mean number of baryons and antibaryons and they are
set to unity
in the final formulae.
By expressing  $\delta_K$ as
$$
\delta_K(n)=\frac{1}{2\pi}\int_0^{2\pi}d\phi~ e^{-i n\phi}~,
$$
Eq.~(\ref{partfunc}) becomes 
\begin{eqnarray}\label{partfunc1}
Z(T,V,B)~&=&~\frac{1}{2\pi}\int_0^{2\pi}d\phi~ e^{-i B\phi} 
~\prod _{j}\sum_{N_b^{(j)}=0}^{\infty}\sum_{N_a^{(j)}=0}^{\infty}       
\frac{(\lambda_b^{(j)} z_j~e^{i\phi})^{N_b^{(j)}}}{N_b^{(j)}!}
\frac{(\lambda_a^{(j)} z_j~e^{-i\phi})^{N_a^{(j)}}}{N_a^{(j)}!}~
=~\nonumber \\
&=&~\frac{1}{2\pi}\int_0^{2\pi}d\phi~ e^{-i B\phi}~ 
\exp\left[\sum _j z_j(\lambda_b^{(j)}~e^{i\phi}~+~ 
\lambda_a^{(j)}~e^{-i\phi})\right]~. 
\end{eqnarray}
This form of the $c.e.$ partition function allows one
to derive
the mean numbers of baryons and antibaryons
\begin{equation}\label{np}
\langle N_b^{(j)} \rangle ~=~ \left(\frac{\partial \log Z}{\partial 
\lambda_b^{(j)}}
\right)_{\lambda_b=\lambda_a=1}~=~z_j~\frac{Z(T,V,B-1)}{Z(T,V,B)}~,
\end{equation}
\begin{equation}\label{na}
\langle N_a^{(j)} \rangle ~=~ \left(\frac{\partial \log Z}{\partial 
\lambda_a^{(j)}} \right)_{\lambda_p=\lambda_a=1}~=~
z_j~\frac{Z(T,V,B+1)}{Z(T,V,B)}~.
\end{equation}
For $\lambda_b=\lambda_a=1$ the partition
function (\ref{partfunc1}) can be presented 
as the modified Bessel function
\begin{equation}\label{partfunc2} 
Z(T,V,B)~=~\frac{1}{2\pi}\int_0^{2\pi}d\phi~ e^{-i B\phi}~
\exp(2z~\cos\phi)~=~I_B(2z)~,
\end{equation}
where $z\equiv \sum_j z_j$. 
This yields final expressions for the mean number of baryons and antibaryons 
\begin{equation}\label{np1}
\langle N_b^{(j)} \rangle ~=~ z_j ~\frac{I_{B-1}(2z)}{I_{B}(2z)}~,~~~~
\langle N_a^{(j)} \rangle ~=~ z_j ~\frac{I_{B+1}(2z)}{I_{B}(2z)}~.
\end{equation}
As the exact baryon number conservation is imposed on
each microscopic state it is evidently fulfilled also by the average values
(\ref{np1}):
\begin{equation}\label{cons}
\langle N_b \rangle ~ - ~ \langle N_a \rangle ~
\equiv ~ \sum_j \langle N_b^{(j)} \rangle ~ -~ 
\sum_j \langle N_a^{(j)} \rangle ~ =~  B~, 
\end{equation}
as indeed can be easily seen from the identity
$
I_{n-1}(x)-I_{n+1}(x)=2nI_n(x)/x
$
\cite{I}.  
Eq.~(\ref{np1}) is valid for all 
combinations of $B$ and $z$ values. 
For a specific case of $B=0$ in the nucleon--antinucleon gas (i.e., no
resonances included) our results (\ref{np1}) are reduced to the result
of Rafelski and Danos \cite{Ra:80}.

The $c.e.$ expressions for the mean number of baryons and
antibaryons can be further simplified for the two
limiting cases: $z\ll 1$ (small systems) and
$z\gg1$ (large systems).
Using the representation of $I_n$ as
the infinite series \cite{I}
$$
I_n(2z)~ = ~ \sum_{k=0}^{\infty} \frac{z^{n+2k}}{k!(n+k)!}~,
$$
 one obtains
for small 
systems
\begin{equation}\label{np2}
\langle N_b^{(j)} \rangle
~\cong~ B~\frac{z_j}{z}+~\frac{z_j\cdot z}{B+1}~+~o(z_j\cdot
z^3)~,~~~~
\langle N_a^{(j)} \rangle
~\cong~ \frac{z_j\cdot z}{B+1}~+~o(z_j\cdot z^3)~.
\end{equation}
The dependence $\langle N_a \rangle \propto V^2/(B+1)$
is therefore observed from Eq.~(\ref{np2}) for the antibaryon
yield in small systems.
Such a dependence
can be intuitively understood from the kinetic picture 
of the baryon--antibaryon pair creation and annihilation.
Let's consider the time dependence $N_a(t)$
(the time averaging of $N_a(t)$ should reproduce our statistical  
average $\langle N_a \rangle$). At each time moment the value of
$N_a(t)$ equals 0 or 1
(configurations with $N_a(t)\ge 2$ can be safely neglected
as $\langle N_a \rangle \ll 1$).
The kinetics of the evolution of $N_a(t)$
is defined by the frequency
$\omega$ of the baryon--antibaryon pair creation and
the life--time $\Delta t$ of
the produced
baryon--antibaryon pair:
$\langle N_a(t) \rangle =\omega \Delta t.$
As a baryon--antibaryon pair is locally produced in any point of the
system
we have $\omega \propto V$. Being produced as baryon--antibaryon 
pair
the antibaryon can be then locally annihilated by any baryon
existing in the system. Therefore, $\Delta t \propto 1/n_B$
($n_B$ is the density of baryons).
$n_B=1/V$ for $B=0$ (as only one baryon exists in the volume $V$),
and $n_B=(B+1)/V$ if $B$ baryons are present in the system
before the baryon--antibaryon pair creation.
These lead to the dependence of the antibaryon number 
on $V$ and $B$ as given by Eq.~(\ref{np2}).

For large systems ($z\gg1$) the {\it c.e.}
becomes equivalent to the {\it g.c.e.}
where the partition function and the average number of baryons 
and antibaryons are calculated as
 \begin{equation}\label{gce}
Z(V,T,\mu_B)~=~\sum_b ~\exp\left(\frac{\mu_B~ b}{T}\right)
~Z(T,V,b)~=~\exp \left(ze^{\mu_B/T}~+~ze^{-\mu_B/T}\right)~,
\end{equation}
\begin{equation}\label{pngc}
\langle N_b^{(j)} \rangle~=~z_j~\exp(\mu_B/T)~,~~~~
\langle N_a^{(j)} \rangle~=~z_j~\exp(-\mu_B/T)~.
\end{equation}
Here $\mu_B$ is a baryon chemical potential
which using Eq.~(\ref{cons}) is defined for $B=\langle b\rangle \ge 0$ as:
\begin{equation}\label{mub}
\exp(\mu_B/T)~=~\frac{B}{2z}~+~\sqrt{1+\left(\frac{B}{2z}\right)^2}~.
\end{equation}
Note that the function $f_j=f_j(T)$ introduced in Eq.~(\ref{zfunc})
has the physical meaning of the density of the $j$-th baryon and
antibaryon in the {\it g.c.e.}
formulation for $\mu_B=0$.
Using the uniform asymptotic expansion of 
the modified Bessel functions at $n\rightarrow \infty$  \cite{I}
\begin{equation}\label{unifasym}
I_n(nx)\cong \frac{1}{\sqrt{2\pi n}}~\frac{\exp(n\eta)}{(1+x^2)^{1/4}}~
\left[1+o\left(\frac{1}{n}\right)\right]~;~~~
\eta\equiv \sqrt{1+x^2}+\ln\frac{x}{1+\sqrt{1+x^2}}~,
\end{equation}
results (\ref{pngc}) and (\ref{mub}) for the 
$g.c.e.$
are also easy to obtain from 
the $c.e$.
(\ref{np1}) in the thermodynamical
limit $V \rightarrow \infty, B\rightarrow \infty$ with
$B/V\equiv \rho_B = const(V)$. 

\vspace{0.1cm}
In order to remove a 'trivial' linear dependence of the 
particle multiplicities on the system volume it is convenient to make a
comparison between the particle ratios from the model and the
experimental data. 
In the statistical model the multiplicity of  
any 'neutral' meson state $M$
is 
just proportional to the volume
($\langle N_M \rangle=z_M=
Vf_M(T)$) for both small ($c.e.$) and large ($g.c.e.$)  systems.
Therefore, the system volume dependence of the
ratio $\langle N_a \rangle/\langle N_M \rangle$
at fixed temperature
is the same as for the  
antibaryon density. From 
Eqs.~(\ref{np1}) and (\ref{pngc}) 
the antibaryon 
densities 
for the {\it c.e.} and {\it g.c.e.} are equal to
\begin{eqnarray}
\frac{\langle N_a^{(j)} \rangle}
{V}\mid_{ce} &\equiv& 
n_a^{(j)}\mid_{ce} = f_j~\frac{I_{B+1}(2z)}{I_{B}(2z)}~
=~f_j~\frac{I_{B+1}(x B)}{I_{B}(x B)}
~\cong~
f_j~ \frac{x}{2}~\frac{B}{B+1}~,\label{adens1} \\
\frac{\langle N_a^{(j)} \rangle} 
{V}\mid_{gce} & \equiv &n_a^{(j)}\mid_{gce}=
 f_j ~\exp(-\mu_B/T)~=~f_j~
\frac{x}{1+\sqrt{1+x^2}}~, \label{adens2}
\end{eqnarray}
where $f\equiv \sum_j f_j$, $x \equiv 2z/B = 2f/\rho_B$, and the last 
approximation in 
Eq.~(\ref{adens1}) is valid for small systems only.
Note that introducing the variable $x$ we have transformed
the finite size $V$-dependence of the $c.e.$ density (\ref{adens1})
into its dependence on the baryon number $B$. 
Eqs.~(\ref{adens1},\ref{adens2}) 
give us the primary thermal density for all individual
antibaryon states $j$. Each non-strange resonance (anti)baryon
state
decays finally into (anti)nucleon plus meson(s).
Therefore, the total (primary plus resonance decay)
antinucleon density equals to the total thermal antibaryon density, 
$n_a=\sum_jn_a^{(j)}$ and is given by  
 Eqs.~(\ref{adens1},\ref{adens2}) with the substitution of
$f_j$ by a sum $f=\sum_j f_j$.

For the purpose of the following discussion
we define a canonical suppression factor
\begin{equation}\label{cs}
F_{cs}~\equiv~\frac {\left(n_a\right)_{ce}}
{\left(n_a\right)_{gce}}~.
\end{equation}
It quantifies the antinucleon suppression due to the exact
baryon number conservation. We note also that the suppression
factor $F_{cs}$ (\ref{cs}) is the same for any individual
antibaryon state.

\vspace{0.5cm}
\noindent
{\bf 3. Discussion}

The results derived in the previous section are used
here to discuss antibaryon production in 
high energy collisions.

In the $B = 0$ case 
the baryon and antibaryon densities are
equal and
Eqs.~(\ref{adens1}) and (\ref{adens2}) yield
\begin{equation}\label{bzero}
n_a^{(j)}\mid_{ce}  = 
n_b^{(j)}\mid_{ce} =
~f_j~\frac{I_{1}(2z)}{I_{0}(2z)}~\cong~f_j~z,~~~
n_a^{(j)}\mid_{gce}=
n_b^{(j)}\mid_{gce}= ~f_j~,
\end{equation}
where the approximation for the 
{\it c.e.} density is valid for small system only.
The canonical suppression factor (\ref{cs}) for $B=0$ 
 is equal to
\begin{equation}\label{cs0}
F_{cs}^{0}~=~\frac{I_{1}(2z)}{I_{0}(2z)}~\cong~f~V~.
\end{equation}
The approximation in
Eq.~(\ref{cs0}) is valid for small system only ($z\equiv fV\ll 1$).

The behavior of the canonical suppression factor $F_{cs}^{0}$
(\ref{cs0}) is shown by the solid lines in Fig.~1 for $T=160$~MeV,
170~MeV and 180~MeV,
assuming that $f$ is the sum of $f_j$ over
all non-strange baryons.
The lines start from $V = $ 5~fm$^3$,
which is approximately equal to the estimate of 
the hadronization volume
for $e^++e^-$ interactions at $\sqrt{s}=29$~GeV \cite{Be:96}.
One observes (see Fig.~1) a strong $c.e.$ suppression 
of the (anti)baryon density.
For $T=160$~MeV the (anti)baryon density increases by a factor of 10 
from its value at $V=5~$fm$^3$ to its $V\rightarrow \infty$
{\it g.c.e.} limit. 
For the small systems   
the (anti)baryon density increases
approximately linearly with $V$, i.e., the (anti)baryon 
multiplicity for the small systems is proportional to $V^2$. 
The $c.e.$ suppression becomes less pronounced and
the volume region with linear increase of the 
(anti)baryon density is reduced for increasing temperature.

Let us now turn to the antibaryon production
in baryon rich system.
In the analysis of data on particle multiplicities
in p+p, p+A and A+A collisions
one usually assumes that all participating nucleons in the collisions
(wounded nucleons) take part in the statistical
hadronization of  the system.
It means that in the analysis 
of the NA49 results 
on antiprotons from p+p interactions to central Pb+Pb collisions
at 158 A$\cdot$GeV we 
should study statistical systems with $2 \le B \le 400$.
It was  found that
the mean multiplicity of pions per wounded nucleon increases
(at the SPS collision energies)
only by about 20\% when going from p+p interactions to central
Pb+Pb collisions \cite{Ga:97}.
The pion to baryon ratio in the statistical model is determined
by two parameters: the temperature and baryon density.
Thus as the temperature is found to be constant 
($T= 175\pm 15$ MeV) we conclude
that the
baryon density at hadronization in nuclear collisions at
158  A$\cdot$GeV is also approximately constant.

Therefore, for the comparison with the NA49 results
we study the evolution of the antibaryon density
with increasing net baryon number $B$ at 
$T = const$ and $\rho_B = const$.
The $c.e.$ suppression factor (\ref{cs}) is found
at these conditions from 
Eqs.~(\ref{adens1},\ref{adens2})
\begin{equation}\label{csb}
F^B_{cs}~=~\frac{1+\sqrt{1+x^2}}{x}~\frac{I_{B+1}(xB)}{I_{B}(xB)}~;~~~~~~~~~~
x\equiv \frac{2f}{\rho_B}~.
\end{equation}
Its $B$-dependence 
is plotted in Fig.~2  for several different values of the parameter $x$.
Note that our assumption $T = const$ and $\rho_B = const$
for  statistical hadronization at different values of $B$ can be 
substituted by a weaker one, $x = const$. From Fig.~2 
one observes that the $c.e.$ suppression of antibaryon density 
becomes stronger at high baryon density (i.e., small $x$).
For $x<1$ the $c.e.$ suppression $F^B_{cs}$ (\ref{csb}) becomes 
close to its $x\rightarrow 0$ limit:
\begin{equation}\label{cslim}
F^B_{cs}~=~\frac{B}{B+1}~.
\end{equation}
Eq.~(\ref{cslim}) shows that 
the strongest $c.e.$ suppression of the antibaryon 
density is for the $B=2$ (nucleon--nucleon interactions) case and it 
leads to
the suppression factor of 2/3.  
This moderate effect of $c.e.$ suppression is in  strong contrast with 
the large $c.e.$ suppression (i.e., $F^0_{cs}\ll 1$)
in the baryon--free system.
A  mathematical reason of this very different behavior
for $B=0$ and $B\ge 2$ (with $\rho_B = const(V)$)
is due to the fact that in the latter case 
both the order of the modified Bessel functions and their arguments
are dependent on $B$ (i.e., on $V$) whereas in the $B=0$ case
only the argument increases with $V$. 

The presence of non-zero baryon number $B > 0$ has a twofold effect
on antibaryon production. First, it suppresses 
the production of antibaryons:
the additional factors $\exp(- \mu_B/T)=x/(1+\sqrt{1+x^2}) < 1$ and 
$1/(B+1) < 1$
appear respectively in the 'large' and 'small' systems    
for the antibaryon density in  comparison with the $B=0$ case. On the
other hand,
the $c.e.$ suppression effect due to 
the exact baryon number conservation
becomes smaller: at fixed $T$ and $V$ the following inequality
is always valid, $F_{cs}^B > F_{cs}^0$. 
For fixed $B>0$ the $c.e.$ suppression of antibaryons
becomes smaller when $\rho_B$ decreases 
and it disappears completely 
(i.e., $F_{cs}^B \rightarrow 1$) in the limit 
 $\rho_B \rightarrow 0$ (and respectively 
 $V \rightarrow \infty$ in order to keep the $B$ value fixed).
This is because the total number 
of baryon--antibaryon pairs becomes large due to large $V$.
Note that in this case the last approximation in Eq.~(\ref{adens1})
is no more valid. Instead one should use the large argument asymptotic
of the modified Bessel functions.

Thus for $B \ge 2$ systems at constant $x=2f/\rho_B$ 
the $c.e.$ suppression 
factor $F^B_{cs}$ (\ref{csb}) ranges between 2/3 
and 1 for $x \ll 1$ and between $(1 - 1/4x)$ and 1 for $x \gg 1$.

Previous analyses of  hadron production at the CERN SPS
indicate large baryon densities at hadronization.
The typical values of $T\cong 170$~MeV and $\mu_B \cong 
250$~MeV found for Pb+Pb collisions \cite{Go:99} lead to the estimate  
$x=2f/\rho_B =sh^{-1}(\mu_B/T)\approx 0.5$. 
As seen from Fig.~2 the $c.e.$ 
suppression factor $F^B_{cs}$ (\ref{csb})
for this 'small' value of $x$ is close to its limiting pattern
$B/(B+1)$ (\ref{cslim}).
In Fig.~3 the NA49 results \cite{Ve:99}
on the $\bar{p}/\pi$ ratio in p+p and Pb+Pb collisions
at 158 A$\cdot$GeV  are compared 
with this limiting pattern. From 
this comparison we conclude that the model of statistical
production of antiprotons at hadronization in baryon--rich
system correctly reproduces the observed {\it antiproton scaling}.

Let us return again to the case of the baryon--free system.
The statistical model calculations 
for e$^+$+e$^-$ \cite{Be:96} and p+$\bar{\rm{p}}$ \cite{Be:97}
interactions include 
 large $c.e.$ suppression
effects. 
As discussed in the introduction
we assume that the hadronization
temperature reflects a universal property of the hadronization
process and therefore should be collision energy independent.
In the case of e$^+$+e$^-$ \cite{Be:96} interactions the hadronization
volume is small and therefore one expects
approximate proportionality to $V^2$  
of the multiplicity of nucleon--antinucleon 
pairs but only 
a linear increase with $V$ of the pion multiplicity.
Therefore, the $\bar{p}/\pi$ ratio ratio calculated within the model 
increases linearly with increasing pion multiplicity.
However experimental data \cite{pbar, pi} contradict this expectation
of the statistical model.
The $\bar{p}/\pi$ ratio which is plotted in Fig. 4
as a function of pion multiplicity 
for e$^+$+e$^-$ interactions at different energies,
$\sqrt{s}=14\div 91$~GeV,
is approximately  constant.

Within the discussed statistical
hadronization model one can try to  
solve the problem by assuming an increase of the temperature $T$  
with decreasing volume $V$.
The function $f(T)$  strongly increases with $T$ which allows
to compensate the $c.e.$ suppression effect 
(to keep the (anti)nucleon density, $ n_a \cong f^2(T)V$,  
constant)  for moderate (of about 10 MeV) changes of $T$. 
This  indeed is observed in the fit results of
the statistical model for the e$^+$+e$^-$ and 
p+$\bar{\rm{p}}$ data:
the increase of the volume is always
accompanied with the decrease of the temperature parameter 
\cite{Be:96,Be:97}.  
Thus one may argue
that the hadronization
condition $T=const$ 
has to be substituted by a different
criterion which should explain the  decrease  of
the temperature with increasing size of the system
in e$^+$+e$^-$ and p+$\bar{\rm{p}}$ interactions. 
The constant energy per particle was recently discussed  as a
chemical freeze--out condition \cite{Re:98}.
The detailed study of this question is, however,
outside of the scope of the present paper.
We note only that the statistical production of 
heavy particles (e.g., $J/\psi$ mesons \cite{Ga:99})
is very sensitive to the temperature parameter. Their yields,
therefore, can be used to clarify the problem.

\vspace{0.5cm}
\noindent
{\bf 4. Summary}

The role of baryon number conservation 
in the calculations 
of antibaryon multiplicity within statistical model of hadronization
was investigated.
We derived explicit analytical formulae for the antibaryon
multiplicity in baryon--free and baryon--rich small and large
systems.
This formalism was further used to discuss {\it antiproton scaling}
observed experimentally in A+A 
collisions.
The statistical model with constant hadronization temperature correctly
reproduces the weak dependence
of the $\bar{p}/\pi$ ratio on the system size in 
p+p and nuclear collisions at the CERN SPS energy.
A description of the ratio of $J/\psi$ mesons to
pions within the statistical hadronization model requires also
a constant temperature parameter in p+p and A+A
collisions at the CERN SPS.
However, the same model with $T= const$ does not give a natural
explanation of the 
approximate independence of the $\bar{p}/\pi$ ratio of collision
energy  in
e$^+$+e$^-$ interactions.
Therefore, a consistent description of  hadron production
within the statistical hadronization model
has not yet been achieved.

\vspace{1cm}
\noindent
{\bf Acknowledgements}

We thank F.~Becattini, K.~Bugaev, J.~Cleymans,
A.~Korol, A.~Kostyuk,
I.~Mishustin, St.~Mr\'owczy\'nski, L.~Neise
P.~Seyboth and H.~St\"ocker for 
fruitful discussions. We acknowledge the
financial support of BMBF and  DFG, Germany.

\newpage
\begin{figure}[p]
\epsfig{file=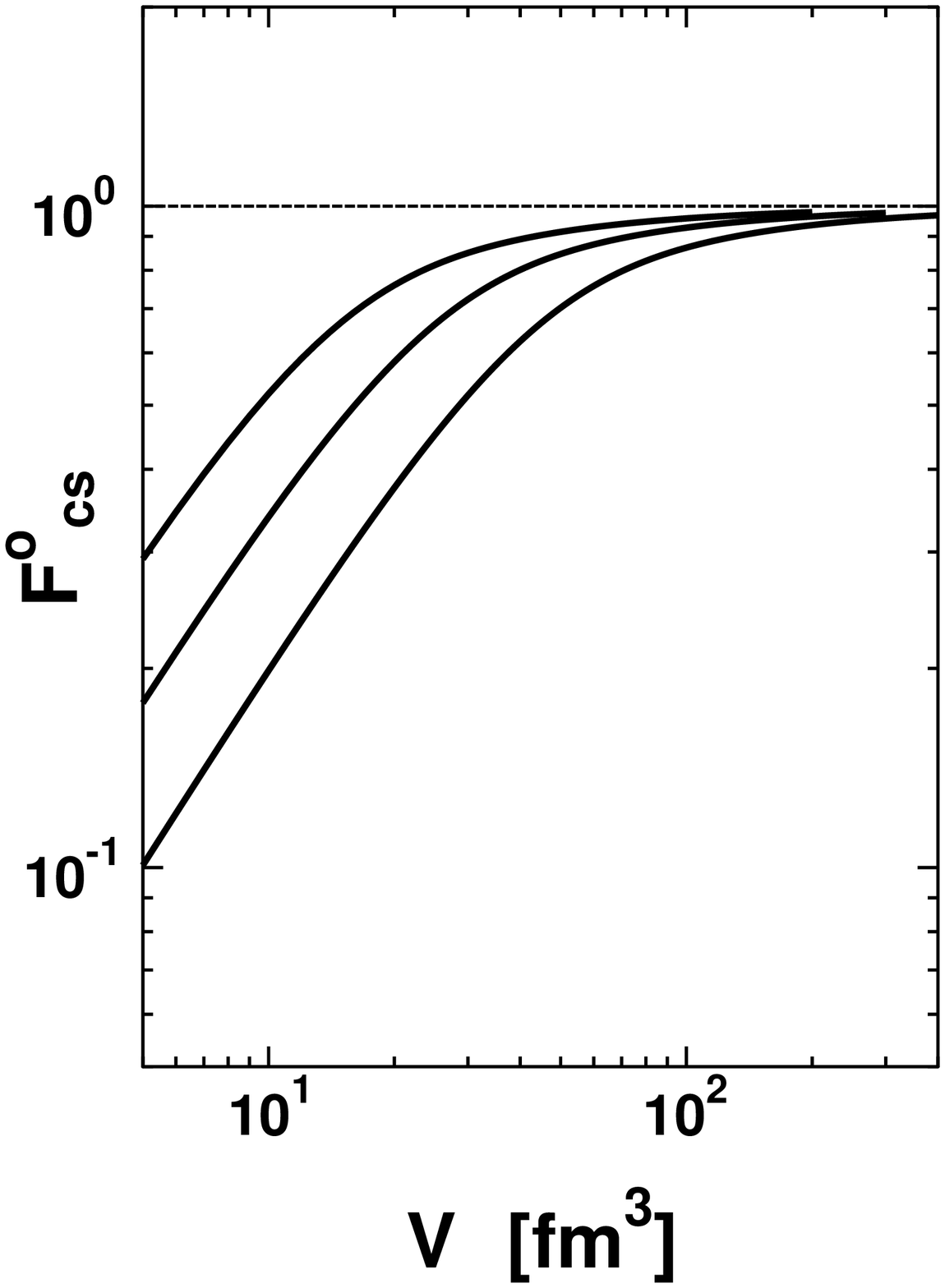,width=10cm}
\caption{
The solid lines show the $c.e.$ suppression factor $F^0_{cs}$ 
(\protect\ref{cs})
for $T=160$~MeV, 170~MeV and 180~MeV (from bottom to top)
for $B = 0$.  
}
\label{fig1}
\end{figure}

\newpage
\begin{figure}[p]
\epsfig{file=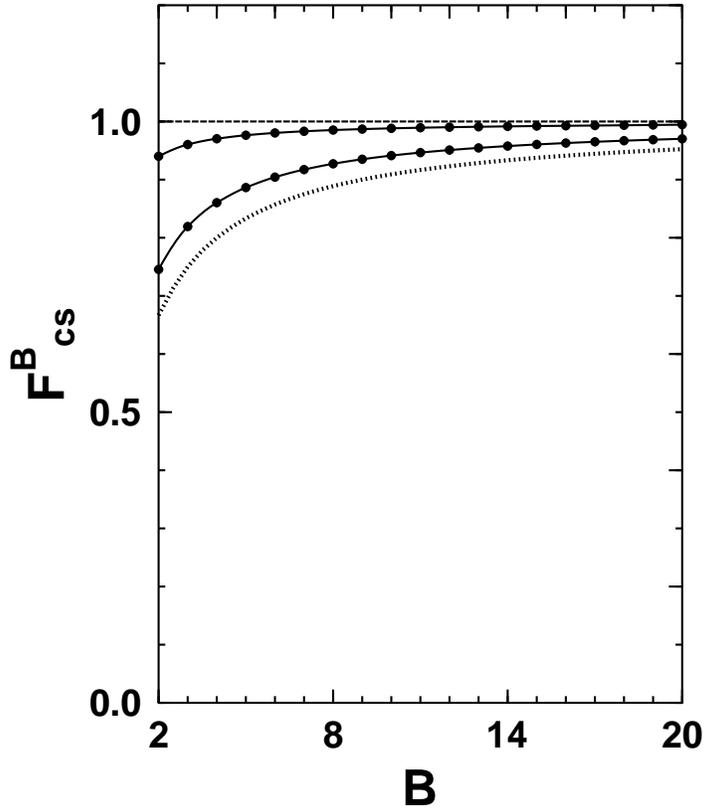,width=10cm}
\caption{
The finite size $B$-dependence of the antibaryon production
in baryon rich ($B\ge 2$) systems at different values
of the variable $x$ ($x \equiv 2z/B = 2f/\rho_B$). 
The solid lines show the $c.e.$ suppression factor $F^B_{cs}$ (\ref{csb})
for $x$=1 and $x$=5 (from below to above). The lower dotted line
corresponds to the limiting $B/(B+1)$ behavior (\ref{cslim}).
}
\label{fig2}
\end{figure}

\newpage
\begin{figure}[p]
\epsfig{file=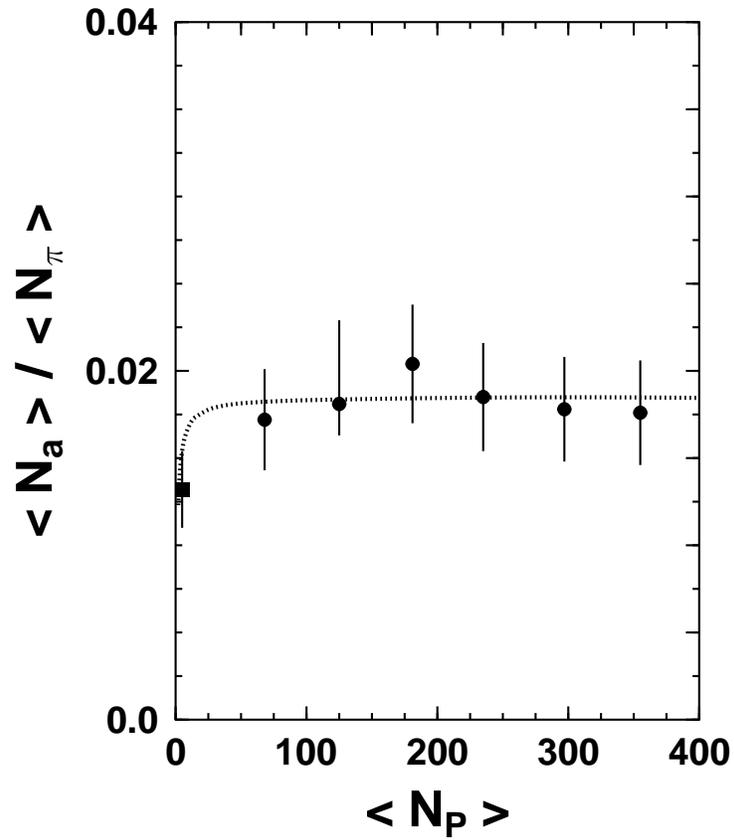,width=10cm}
\caption{
The NA49 data \protect{\cite{Ve:99}}
on the antiproton to pion ratio 
($\langle N_{\pi} \rangle  \equiv ( \langle N_{\pi^+} \rangle +
\langle  N_{\pi^-} \rangle)/2$) 
in p+p (square) and centrality selected Pb+Pb (dots)
collisions at 158 A$\cdot$GeV are plotted as a function
of the mean number of wounded nucleons, $\langle N_P \rangle$.
The dependence of the ratio on $\langle N_P \rangle$  expected
within the statistical model, $\langle N_P \rangle/(\langle N_P
\rangle+1$),
is shown by dotted line, the function is normalized to the 
experimental data.
}
\label{fig3}
\end{figure}

\newpage
\begin{figure}[p]
\epsfig{file=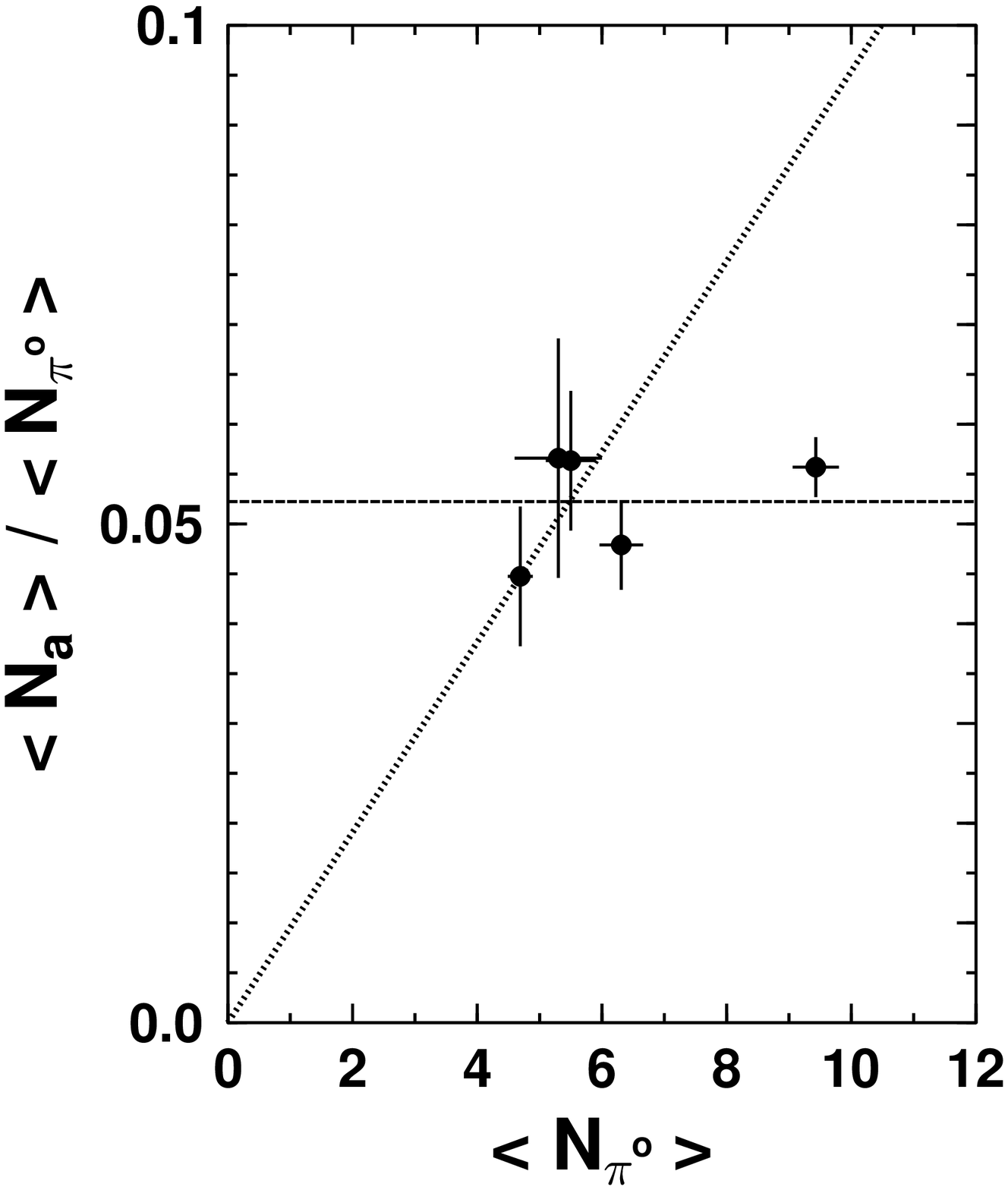,width=10cm}
\caption{
The antiproton
\protect{\cite{pbar}} to $\pi^o$ \protect{\cite{pi}} 
ratio in $e^+e^-$ annihilation at $\sqrt{s}$ = 14, 22, 29 35 and 91~GeV
is plotted as a function of mean multiplicity of $\pi^o$
mesons, $\langle N_{\pi^o} \rangle$.
In order to increase  the significance of the data
the antiproton multiplicity was calculated as 
average of proton and antiproton multiplicities.
The dotted line indicates the proportional
dependence of the ratio on $\langle N_{\pi^o} \rangle$
expected within statistical model for small systems.
The horizontal dashed line is shown for the reference.
}
\label{fig4}
\end{figure}


\begin{thebibliography}{99}

\bibitem{Fe:50} 
E. Fermi, Prog. Theor. Phys. {\bf 5} (1950) 570.

\bibitem{Po:51} 
I. Ya. Pomeranchuk, Dokl. Akad. Nauk SSSR
{\bf 78} (1951) 884.

\bibitem{La:53} 
L. D. Landau, Izv. Akad. Nauk SSSR, Ser. Fiz.
{\bf 17} (1953) 51. 

\bibitem{Go:99}
J. Cleymans and H. Satz, Z. Phys. {\bf C57} (1993) 135.\\
J. Sollfrank, M. Ga\'zdzicki, U. Heinz and J. Rafelski,
Z. Phys. {\bf C61} (1994) 659; \\   
G. D. Yen, M. I. Gorenstein, W. Greiner, S.N. Yang,  
Phys. Rev. {\bf C56} (1997) 2210;\\
F. Becattini, M. Ga\'zdzicki and J. Sollfrank, Eur. Phys. J.
{\bf C5} (1998) 143;\\
G. D. Yen and M. I. Gorenstein,
Phys. Rev. {\bf C59} (1999) 2788;\\
P. Braun--Munzinger, I. Heppe and J. Stachel, Phys. Lett. {\bf 465B}
(1999) 15.

\bibitem{Be:96}
F. Becattini, Z. Phys. {\bf C69} (1996) 485.

\bibitem{Be:97}
F. Becattini and U. Heinz, Z. Phys. {\bf C76} (1997) 269.

\bibitem{temp}
F. Becattini, M. Ga\'zdzicki and J. Sollfrank,
Nucl. Phys. {\bf A638} (1998) 403.


\bibitem{Ke:99}
A. Ker\"anen, J. Cleymans and E. Suhonen,
J. Phys. {\bf G25} (1999) 275.

\bibitem{Ve:99}
G. Veres et al. (NA49 Collab.), 
Proceedings of the Forthteens International Conference
on Ultra--Relativistic Nucleus--Nucleus Collisions,
Torino, Italy, May 1999.

\bibitem{Ha:71}
R. Hagedorn, CERN yellow report No. 71-12, 1971.

\bibitem{Re:80}
K. Redlich and L. Turko, Z. Phys. {\bf C5} (1980) 541.

\bibitem{Ra:80}
J. Rafelski and M. Danos, Phys. Lett. {\bf B97} (1980) 279.

\bibitem{Cl:97}
J. Cleymans, A. Ker\"anen, M. Marais and E. Suhonen,
Phys. Rev. {\bf C56} (1997) 2747.


\bibitem{pbar}
M. Althoff et al. (TASSO Collab.), Z. Phys. {\bf C27}
(1985) 27, \\
H. Aihara et al. (TPC Collab.), Phys. Rev. Lett.
{\bf 52} (1884) 577, \\
H. Aihara et al. (TPC Collab.),
Phys. Lett. {\bf 184B} (1987) 299, \\
W. Bartel et al. (JADE Collab.),
Phys. Lett. {\bf 104B} (1981) 325, \\
W. Braunschweig et al. (TASSO Collab.),
Z. Phys. {\bf C42} (1989) 189, \\
P. Abren et al. (DELPHI Collab.),
Nucl. Phys. {\bf B444} (1995) 3, \\
R. Akers et al. (OPAL Collab.),
Z. Phys. {\bf C63} (1994) 181, \\
P. Abren et al. (DELPHI Collab.),
Eur. Phys. J. {\bf C5} (1998) 585, \\
K. Abe et al. (SLD Collab.),
Phys. Rev. {\bf D59} (1999) 052001.

\bibitem{pi}
W. Bartel et al. (JADE Collab.),
Z. Phys. {\bf C28} (1985) 343, \\
H. Aihara et al. (TPC Collab.),
Z. Phys. {\bf C27} (1985) 187, \\
W. Braunschweig et al. (TASSO Collab.),
Z. Phys.  {\bf C33} (1985) 13, \\
H. J. Behrend et al. (CELLO Collab.),
Z. Phys. {\bf C47} (1990) 1, \\
D. Pitzl et al. (JADE Collab.),
Z. Phys. {\bf C46} (1990) 1, \\
M. Acciarri et al. (L3 Collab.),
Phys. Lett. {\bf 328B} (1994) 223, \\
W, Adam et al. (DELPHI Collab.),
Z. Phys. {\bf C69} (1996) 561, \\

\bibitem{I} 
M. Abramowitz  and I.E. Stegun,  
 Handbook of Mathematical Functions, 1964 (New York: Dover).

\bibitem{Ga:97}
M. Ga\'zdzicki, J. Phys. {\bf G23} (1997) 1881.

\bibitem{Re:98}
J. Cleymans and K. Redlich, Phys. Rev. Lett. 
{\bf 81} (1998) 5284.

\bibitem{Ga:99}
M. Ga\'zdzicki and M. I. Gorenstein,
Phys. Rev. Lett.
{\bf 83} (1999) 4009.
 
\end{thebibliography}
\end{document}